\documentclass[letterpaper,12pt]{article}

\title{Gravitational Friedel oscillations in\\higher-derivative and infinite-derivative gravity?}

\author{Jens Boos\,\footnote{~E-mail: \href{mailto:boos@ualberta.ca}{boos@ualberta.ca}} \\
    {\small Theoretical Physics Institute, University of Alberta, Edmonton, AB T6G 2E1, Canada} }

\date{Oct 19, 2018}

\usepackage{epsfig}
\usepackage[english]{babel}
\usepackage{amsmath}
\usepackage{amsbsy}
\usepackage{amstext}
\usepackage{amscd}
\usepackage{amsxtra}
\usepackage{amsopn}
\usepackage[onehalfspacing]{setspace}
\usepackage{tensor}
\usepackage{mathrsfs}
\usepackage{relsize}
\usepackage{amsfonts}
\usepackage[makeroom]{cancel}
\usepackage{selinput}

\usepackage{graphics}
\usepackage[export]{adjustbox}
\usepackage{subfig}

\usepackage{youngtab}

\newcommand{\dd}{\mbox{d}}

\newcommand{\lap}{\bigtriangleup}
\newcommand{\ts}[1]{{\boldsymbol{#1}}}

\newcommand{\ind}[1]{{\mbox{\scriptsize #1}}}

\usepackage{geometry}
\usepackage{parskip}

\usepackage[colorlinks=true,citecolor=green,linkcolor=red,filecolor=cyan,urlcolor=magenta]{hyperref}

\usepackage{color}
\definecolor{mygray}{rgb}{0.5,0.5,0.5}

\usepackage{listings} \definecolor{fadedred}{rgb}{0.6,0.0,0.0}
\definecolor{fadedblue}{rgb}{0.0,0.0,0.6}

\begin{document}

\maketitle

\begin{abstract}
When a positively charged impurity is placed inside a cold metal, the resulting charge density around that object exhibits characteristic ripples to negative values, known as \emph{Friedel oscillations}. In this essay, we describe a somewhat analogous effect in (i) linearized higher-derivative gravity and (ii) linearized infinite-derivative ``ghost-free'' gravity: when a gravitational impurity (point particle) is placed in Minkowski vacuum, the local energy density $\rho \equiv G{}_{\mu\nu}\xi{}^\mu \xi{}^\nu$ (with $\xi=\partial_t$) exhibits oscillations to negative values. The wavelength of these oscillations is roughly given by (i) the Pauli--Villars regularization scale and (ii) the scale of non-locality. We hence dub this phenomenon \emph{gravitational Friedel oscillations}.

{\hfill \smaller \textit{file: gravitational-friedel-oscillations-v6.tex, Oct 19, 2018, jb} }

\vspace{6cm} {\smaller Essay written for the Gravity Research Foundation 2018 Awards for Essays on Gravitation.}

\end{abstract}

\vfill

\pagebreak


\section{Introduction}
It is well known that Einstein's General Relativity (GR) is plagued by short-scale divergences, be it in the context of the curvature singularities inside black holes, or the big bang and related singularities in cosmology. In the context of point masses, the the gravitational field close to the mass becomes singular, and curvature invariants diverge. These singularities ``survive'' the Newtonian limit, where they resurface as unbounded tidal forces. Seemingly, general covariance is not the guiding principle to ameliorate the situation. On the other hand, these singularities are unphysical and have to be dealt with. What to do?

One may hope that these issues will be resolved once a consistent UV completion of GR (``quantum gravity'') is known. In the meantime, we can instead attempt to find modifications of GR that feature finite potentials, regarding these modifications as the effective field theory limit of some more fundamental concepts, hitherto unknown. Before jumping into technical details, let us consider some modifications of Newtonian gravity by considering the field of a $\delta$-like mass distribution
\begin{align}
\label{eq:delta-density}
\rho = 4\pi G m\delta(r) \, .
\end{align}
Then, the Poisson equation gives the well-known Newtonian potential of a point mass,
\begin{align}
\lap \phi = \rho \, , \quad \phi(r) = -\frac{Gm}{r} \, .
\end{align}
This potential is singular, and it gives rise to infinite tidal forces. In this essay, we shall uphold Eq.~\eqref{eq:delta-density}, that is, the notion of sharply concentrated densities. Then, a simple Pauli--Villars-type regularization scheme of the Poisson equation can ameliorate the situation:
\begin{align}
\label{eq:modified-poisson}
\lap (1 + M^{-2}\lap)\phi = \rho \, , \quad \phi(r) = -\frac{Gm}{r} \left( 1 - e^{-M r} \right)\, ,
\end{align}
and $M$ is a large mass scale such that for $M \rightarrow \infty$ one recovers the original Poisson equation. For finite $M$, however, the potential is now finite at the origin, $\phi(0) = -GMm$. Note, however, that it is not regular since its derivative does not vanish: $\phi'(0) = GmM^2/2$. If we think of the Newtonian limit of GR, this would imply that the corresponding spacetime has a conical singularity at the origin. There is another problem, however, which is generic to higher-derivative modifications:

Typically, they bring along massive ghost degrees of freedom, which in turn lead to unstable vacua upon quantization. In the above example, the Green function is
\begin{align}
D(r) = \frac{1}{\lap(1 + M^{-2}\lap)} = \frac{1}{\lap} - \frac{M^2}{M^2 + \lap} \, ,
\end{align}
the second term of which corresponding to a ghost of mass $M$. The particle spectrum of this theory, then, will not only feature a massless graviton of helicity 2, but also a massive ghost mode \cite{VanNieuwenhuizen:1973fi,Stelle:1977ry}.

A recent approach \cite{Modesto:2011kw,Biswas:2011ar} that avoids the excitation of ghost modes at tree level \cite{Shapiro:2015uxa} is aptly called \emph{ghost-free gravity}. To see how this works, let us consider the following modification of the Poisson equation:
\begin{align}
e^{f(\lap)} \lap \phi = \rho \, ,
\end{align}
where $f(\lap)$ is some polynomial of the Laplace operator defined as a formal power series (we are forced to introduce again at least one scale $M$ such that we can build the dimensionless combination $M^{-2}\lap$ that enters the power series). Then, the propagator is
\begin{align}
D(r) = \frac{1}{e^{f(\lap)} \lap} \, , 
\end{align}
but by construction there are no new poles since the exponential function is nowhere zero on the real axis. This can be made more rigorous in a mathematical sense by noticing that the exponential of a polynomial is a so-called \emph{entire function} which does not have any poles at finite distance from the origin. Then, one can apply Picard's little theorem for entire functions to ensure that the denominator of the propagator never goes to zero (except at the graviton pole), which is a fundamental mathematical statement that surpasses simple plausibility arguments.

As it turns out (and as we shall see below), the gravitational potential for some generic choices of the function $f(\lap)$ is regular at the origin. However, as a theory with an infinite number of derivatives, this ghost-free gravity is necessarily non-local at some characteristic scale $\ell \equiv M^{-1}$. What is the nature of this non-locality? In this essay we shall treat it as classical, that is, $\ell \gg \ell_\text{Planck}$. This is consistent with regarding ghost-free gravity as an effective description of gravity, obtained by coarse-graining some underlying, more fundamental quantum description. In that sense, $\ell > 0$ may be regarded as a reminder that ghost-free gravity may have a non-classical origin.

Let us summarize: in both higher-derivative and infinite-derivative theories of gravity, one can attain a regular Newtonian potential that is (i) regular at the origin, and (ii) decays like $\sim 1/r$ for large distances. On the classical side there are various attempts to understand the non-linear regime of non-local gravity, whereas on the quantum side the perturbative structures are still to be fully understood.

In the remainder of this essay, we would like to venture in a somewhat orthogonal direction and study the gravitational field of point sources at mesoscopic distances away from the point source, that is, not directly at the origin, and also not very far away where the gravitational field approaches the standard Newtonian $1/r$-behavior.

\section{A framework for linearized higher-derivative gravity}
Let us now briefly sketch a more rigorous framework that we can employ to study the linearized gravitational field of matter distributions in both higher-derivative and infinite-derivative gravity on a flat Minkowski background. Writing the metric as $g{}_{\mu\nu} = \eta{}_{\mu\nu} + h_{\mu\nu}$, the most general action quadratic in $h_{\mu\nu}$ in $D=d+1$ spacetime dimensions can be written as
\begin{align}
\begin{split}
S &= \frac{1}{2\kappa} \int \dd^D x \Big( \frac12 h^{\mu\nu}\,a(\Box)\Box\,h_{\mu\nu}-h^{\mu\nu}\,a(\Box)\partial_{\mu}\partial_{\alpha}\,h^{\alpha}{}_{\nu} +h^{\mu\nu}\, c(\Box)\partial_{\mu}\partial_{\nu} h \\
&\hspace{75pt}  - \frac12 h\,c(\Box)\Box h 
+ \frac12 h^{\mu\nu}\,\frac{a(\Box)-c(\Box)}{\Box}\partial_{\mu}\partial_{\nu}\partial_{\alpha}\partial_{\beta}\,h^{\alpha\beta}\Big) \, ,
\end{split}
\end{align}
where ``$\Box$'' denotes the d'Alembert operator. The two functions $a(\Box)$ and $c(\Box)$ are non-local \emph{form factors} that satisfy $a(0)=c(0)=1$ in order to reproduce linearized GR at large scales. The above action is indeed the most general one since the Bianchi identities relate the possible choices of functions $f(\Box)$ to just two independent functions.

For the sake of simplicity, let us consider a simple case where $a(\Box)=c(\Box)$. Then, for a stress-energy tensor of a point mass, $T{}_{\mu\nu} = m \delta^t_\mu \delta^t_\nu \delta^{(d)}(\vec{r}\,)$, and the metric ansatz
\begin{align}
\label{eq:metric-ansatz}
\dd s^2 = -\left[1-2(d-2)\phi\right]\dd t^2 + (1+2\phi)\dd \vec{r}^{\,2} \, ,
\end{align}
where $\dd \vec{r}^{\,2} = \dd x_1^2 + \dots + \dd x_d^2$ is the metric of flat space in Euclidean coordinates $x_i$ ($i=1,\dots,d$), one obtains the field equations
\begin{align}
a(\lap)\lap \phi = \frac{\kappa m}{d-1} \delta^{(d)}(\vec{r}\,) \, .
\end{align}
The Green function for this static problem takes the form
\begin{align}
\label{eq:green-function}
D(r) = \frac{1}{(2\pi)^{\frac d2} r^{d-2}} \int\limits_0^\infty \dd \zeta \frac{\zeta^{\frac{d-4}{2}}}{a(-\zeta^2/r^2)} J_{\frac d2 - 1}(\zeta) = \frac{1}{2\pi^2 r} \int\limits_0^\infty \dd \zeta \frac{\sin \zeta}{\zeta} \frac{1}{a(-\zeta^2/r^2)} \, .
\end{align}
where $J_n$ denotes the Bessel function of the first kind, and in the second equality we inserted $d=3$ (which we shall concern ourselves with in what follows). The gravitational potential is then given by
\begin{align}
\label{eq:potential-master}
\phi(r) = -\frac{\kappa m}{d-1} D(r) \, ,
\end{align}
and is easy to see that for $a=1$ one obtains the well-known result $\phi(r) = -Gm/r$ as obtained in the Newtonian limit of GR in four spacetime dimensions ($d=3$).

\section{Gravitational Friedel oscillations}
Given the general solution of the potential, Eq.~\eqref{eq:potential-master}, we can now study it shape in various higher-derivative as well as infinite-derivative theories of gravity. The Green function \eqref{eq:green-function} can either be evaluated analytically or numerically; for the sake of this essay, let us focus on analytical results that can easily be written down.

\begin{figure}[!htb]
\centering
\subfloat[Higher derivative theories for $N=1,2,3$.]
{
    \includegraphics[width=0.5\textwidth]{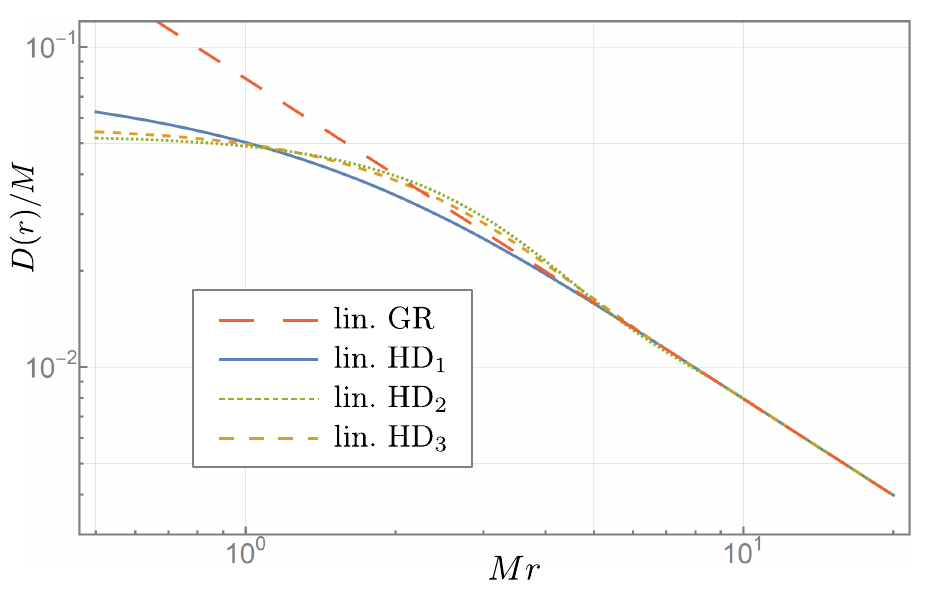}
}
\subfloat[Infinite-derivative theories for $N=1,2,3$.]
{
    \includegraphics[width=0.5\textwidth]{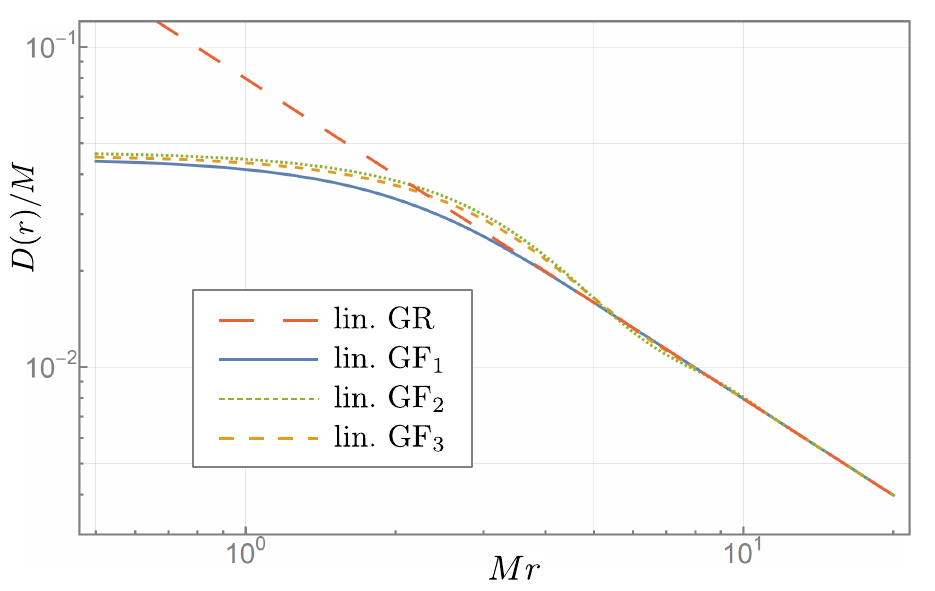}
}
\caption{The Green functions of $\mathrm{HD_N}$ and $\mathrm{GF_N}$ theories visualized for $N=1,2,3$. Whereas $N=1$ approaches the $1/r$ power law directly, there are oscillations in cases of $N=2,3$.}
\label{fig:potentials}
\end{figure}

To that end we shall consider higher-derivative theories of the following class, call them $\mathrm{HD_N}$,
\begin{align}
a(\Box) = 1 + (-\Box/M^2)^N \, , \quad N \in \mathbb{N} \, ,
\end{align}
as well as a class of infinite-derivative ``ghost-free'' theories, call them $\mathrm{GF_N}$:
\begin{align}
a(\Box) = \exp\left[ (-\Box/M^2)^N \right] \, , \quad N \in \mathbb{N} \, .
\end{align}
Clearly these theories satisfy $a=1$ for $M \rightarrow \infty$, so for large scales they will reproduce linearized GR. The Green functions can be calculated analytically, and one obtains (see also \cite{Frolov:2015usa})
\begin{align}
\begin{split}
\label{eq:green-functions}
\mathrm{GR}  : \quad & D(r) = \frac{1}{4\pi r} \, , \\
\mathrm{HD_1}: \quad & D(r) = \frac{1}{4\pi r} \left( 1 - e^{-M r} \right) \, , \\
\mathrm{HD_2}: \quad & D(r) = \frac{1}{4\pi r} \left[ 1 - e^{-Mr/\sqrt{2}} \cos\left( Mr/\sqrt{2} \right) \right] \, , \\
\mathrm{HD_3}: \quad & D(r) = \frac{1}{4\pi r} \left[ 1 - \tfrac13 e^{-Mr} - \tfrac23 e^{-Mr/2} \cos\left( \sqrt{3}Mr/2 \right) \right] \, , \\
\mathrm{GF_1}: \quad & D(r) = \frac{\text{erf}(M r/2)}{4\pi r} \, , \\
\mathrm{GF_2}: \quad & D(r) = \frac{M}{6\pi^2}\Big[ 3 \Gamma\!\left(\tfrac54\right) {}_1\!F\!{}_3\left( \tfrac14;~ \tfrac12,\tfrac34,\tfrac54;~ y^2 \right) -2y\Gamma\!\left(\tfrac34\right) {}_1\!F\!{}_3\left( \tfrac34;~ \tfrac54, \tfrac32, \tfrac74;~ y^2 \right) \Big] \, , \\
\mathrm{GF_3}: \quad & D(r) = \frac{M}{\pi}\Big[ -\frac{1}{\Gamma\left(-\tfrac16\right)}{}_{1\!}F{}_{\!5}\left( \tfrac16;~ \tfrac13, \tfrac 12, \tfrac 23, \tfrac56, \tfrac76;\,-z^3 \right) - \frac{z}{2\sqrt{\pi}} {}_{1\!}F{}_{\!5}\left( \tfrac12;~ \tfrac23,\tfrac56,\tfrac76,\tfrac43,\tfrac32;\,-z^3\right) \\
&\hspace{62pt} + \frac{3z^2}{10\Gamma\left(\tfrac76\right)} {}_{1\!}F{}_{\!5}\left( \tfrac56;~ \tfrac76,\tfrac43,\tfrac32,\tfrac53,\tfrac{11}{6};\,-z^3\right) \Big] \, ,
\end{split}
\end{align}
where ${}_{p\!}F{}_{\!q}(a_1,\dots,a_p;\,b_1,\dots,b_q;\,z)$ denotes the generalized hypergeometric function and we defined the dimensionless radial variables $y \equiv M^2 r^2/16$ and $z \equiv M^2r^2/36$. See Fig.~\ref{fig:potentials} for a visualization of the Green functions: in both higher-derivative and infinite-derivative gravity they are finite at $r=0$, whereas for GR the Green function diverges at the origin.

This constitutes a major insight of these calculations in the literature, and, at the linear level, one can easily extend these studies to $p$-branes in higher-dimensional Minkowski space \cite{Boos:2018bxf}. Observe, however, the particular shape of the Green functions a bit closer. There appears to be a substructure: whereas the $N=1$ Green functions decay like $1/r$ for large values of the dimensionless radial distance $M r$, there exist noticeable oscillations in the potentials for the cases $N=2,3$ \cite{Modesto:2011kw,Modesto:2016ofr,Edholm:2016hbt,Conroy:2017nkc,Boos:2018bxf}; for a visualization, see Fig.~\ref{fig:potentials}.

\begin{figure}[!htb]
\centering
\subfloat[Higher-derivative theories for $N=1,2,3$.]
{
    \includegraphics[width=0.5\textwidth]{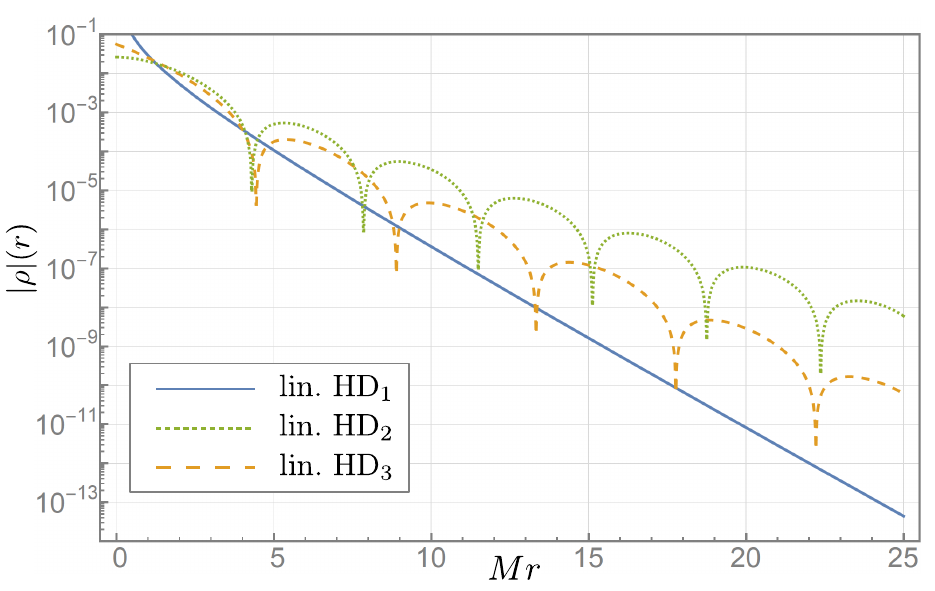}
}
\subfloat[Infinite-derivative theories for $N=1,2,3$.]
{
    \includegraphics[width=0.5\textwidth]{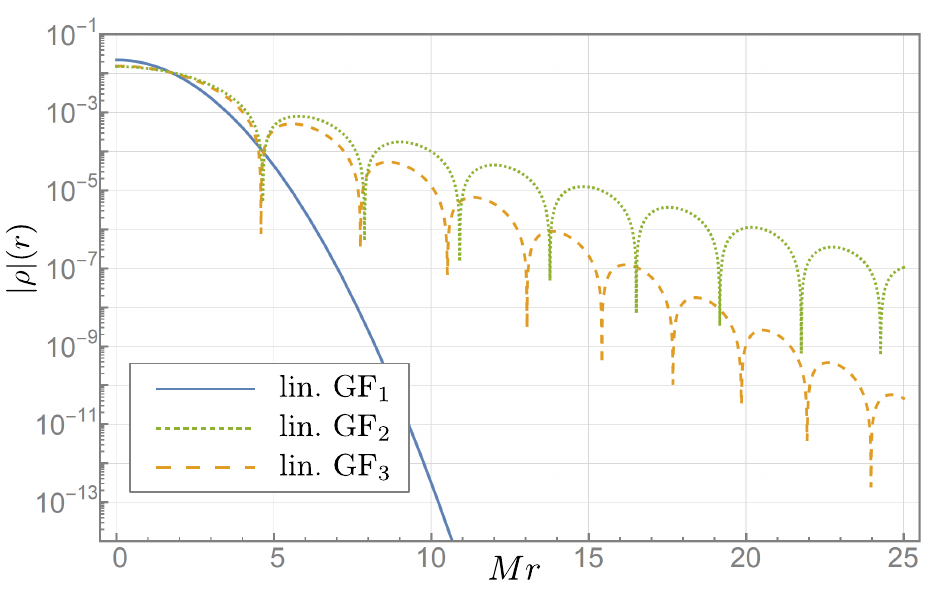}
}
\caption{The absolute value of the local energy density $\rho \equiv G{}_{\mu\nu}\xi{}^\mu\xi{}^\nu$, $\ts{\xi} = \partial_t$, undergoes oscillatory behavior in the cases $N=2,3$, whereas there are no oscillations in the case $N=1$ (both for higher-derivative and infinite-derivative theories). Close to the origin, $M r \approx 0$, one has $\rho > 0$; at the points of diverging slope, the energy density vanishes. Between these points, it switches its overall sign.}
\label{fig:energy-densities}
\end{figure}

These oscillations have direct consequences for the local energy density $\rho$ perceived by a static observer whose 4-velocity is tangential to $\ts{\xi} = \partial_t$. For the metric ansatz \eqref{eq:metric-ansatz} one obtains
\begin{align}
\rho \equiv G{}_{\mu\nu} \xi{}^\mu \xi{}^\nu = (1-d)\lap \phi \, ,
\end{align}
where $G{}_{\mu\nu}$ denotes the linearized Einstein tensor. For $N=1$ theories, this quantity is positive definite, whereas for $N=2,3$ it undergoes oscillations around zero that decay in strength with distance $M r$. See a visualization of this behavior in Fig.~\ref{fig:energy-densities}.

\begin{figure}[!htb]
\centering
\subfloat
{
    \bgroup
	\def\arraystretch{1.2}
	\footnotesize
	\begin{tabular}{rccrcc}
	$M r$ & $M\delta_2/2$ & ~~ & $M r$ & $M\delta_3/2$ \\ \hline
	 4.59 & 3.16 &&  4.65 & 3.23 \\
	 7.75 & 2.76 &&  7.88 & 3.02 \\
	10.51 & 2.53 && 10.90 & 2.86 \\
	13.04 & 2.38 && 13.76 & 2.75 \\
	15.42 & 2.26 && 16.51 & 2.66 \\
	17.68 & 2.17 && 19.17 & 2.58 \\
	19.85 & ---  && 21.75 & ---
	\end{tabular}
	\egroup
}
\qquad
\subfloat{\adjustbox{raise=-6pc}{\includegraphics[width=0.5\textwidth]{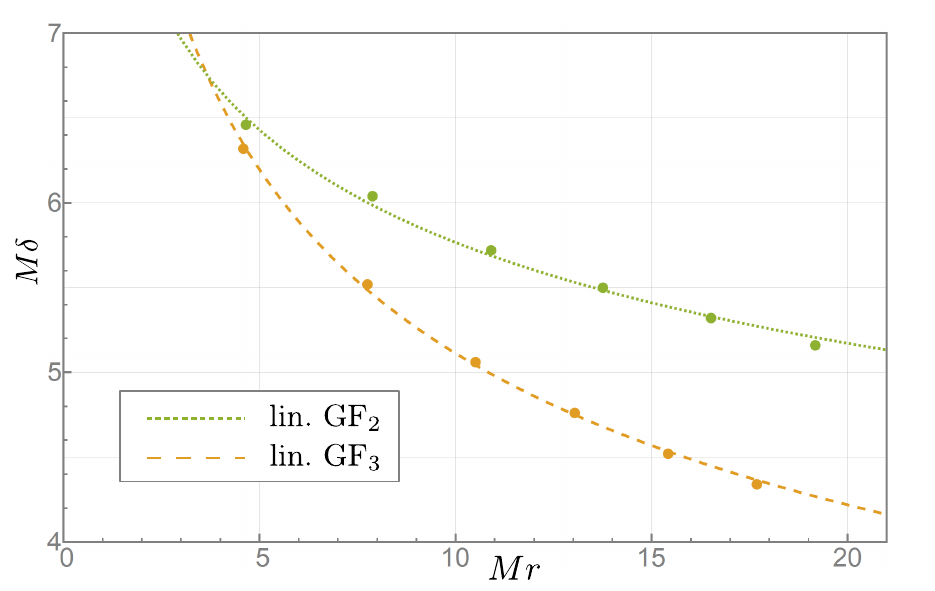}}}
\caption{We evaluated the zeroes of the energy density for $\mathrm{GF_2}$ theory and $\mathrm{GF_3}$ theory numerically, from which we can read off (half the) wavelength $M\delta_N$. Contrary to the higher-derivative theories, in infinite-derivative theories the wavelengths are not constant, but decrease with increasing $M r$. To first approximation, this can be described by simple power laws.}
\label{fig:wavelengths}
\end{figure}

Using these diagrams, we can extract some typical wavelengths: from Eq.~\eqref{eq:green-functions} it is clear that the wavelengths of oscillation are constant in the cases $N=2,3$ in higher-derivative gravity due to the explicit appearance of trigonometric functions. For the infinite-derivative theories the behavior is more involved. The oscillations still scale with $M{}^{-1}$, but they decay with increasing distance $M r$. A rather qualitative fit gives
\begin{align}
\mathrm{GF_2} : \quad M\delta_2 \sim 9.68 \, (M r)^{-0.28} \, , \qquad
\mathrm{GF_3} : \quad M\delta_3 \sim 8.28 \, (M r)^{-0.16} \, ,
\end{align}
but a closer inspection reveals that the precise wavelengths oscillate over and under these curves, see Fig.~\ref{fig:wavelengths} for more details.

It seems that these oscillations occur irrespective of the precise modification method of GR. While they are absent for $N=1$, we should remark that the case $N=1$ is somewhat degenerate for both classes of theories considered: in the higher-derivative framework the potential is indeed finite at the origin, but it's first derivative is non-zero, leading to a conical singularity (as well as a diverging energy density, see Fig.~\ref{fig:energy-densities}). As it turns out, the scalar case of $N=1$ infinite-derivative theory has time-dependent instabilities \cite{Frolov:2016xhq}.

In other words: for all regular versions of both higher-derivative and infinite-derivative gravity, these oscillations do occur at distances where $M r \sim \mathcal{O}(1)$ before they decay roughly like a power law. Since these theories are classical, and hence $M \ll m_\text{Planck}$, the typical distance $r \sim M^{-1} \mathcal{O}(1)$ might be accessible to experiment at some point in time.

Oscillations of energy density, somewhat similar to the ones we described here, are well known in condensed matter physics where they are called \emph{Friedel oscillations} \cite{Friedel:1952,Friedel:1954,Friedel:1958}: upon insertion of a positively charged impurity into a cold metal the overall charge density around this impurity exhibits spatial  oscillations. This effect is usually calculated at 1-loop using the random phase approximation wherein the photon propagator picks up a fermion-loop as a correction term. In other words, the screening mechanism of an electric charge inside a cold metal is to be treated as a scattering problem \cite{Altland:2006}.

There is also a physically intuitive explanation: in the Jellium model, electrons in a metal at low temperature fill up the Fermi sphere up to a maximum momentum of $k_\ind{F}$ while the positive ions form a rigid background structure. Electrons close to the Fermi momentum $k_\ind{F}$ are most prone to interact with the impurity, and since these electrons are non-local objects (scale of non-locality $\sim k_\ind{F}^{-1}$), they cannot compensate the positive charge exactly: they overcompensate the charge, and thereby induce a spatially oscillating charge distribution.

\section{Discussion and conclusions}

In the recent literature, there has been a lot of focus on (i) the classical non-linear behavior and (ii) the perturbative quantum structure of infinite-derivative gravity. In this essay, we pursued a somewhat different direction by focussing on the linearized theory at mesoscopic distances, $M r \sim \mathcal{O}(1)$, where both the gravitational potential as well we the local energy density exhibits fluctuations, the latter one assuming negative values in some regions. For values of $M \gg m_\text{Planck}$ these oscillations might become observable at some point in the future \cite{Accioly:2016qeb,Accioly:2016etf,Perivolaropoulos:2016ucs}.

In an analogy to condensed matter physics and Friedel oscillations in cold metals, we think it may be appropriate to call the oscillations described in this essay \emph{gravitational Friedel oscillations.} Since they do not only appear in higher-derivative theories (wherein they can be interpreted as spurious effects occurring at the Pauli--Villars regularization scale due to the presence of complex poles \cite{Accioly:2016qeb,Giacchini:2016xns}) and they survive the ghost-free limit, we think that these oscillations are of some physical relevance.

At the present stage, the perturbative structure of infinite-derivative ``ghost-free'' gravity is not fully understood \cite{Shapiro:2015uxa,Giacchini:2016xns,Biswas:2013kla,Talaganis:2016ovm} (see, however, the recent work \cite{Calcagni:2018gke,Buoninfante:2018mre}), and it is also not clear whether ghost-ridden higher-derivative theories can be considered physically viable classical theories at distances close to the Pauli--Villars regularizations scale $M$. It would also be interesting to study the linearized gravitational field in other non-local modifications of gravity \cite{Hehl:2009es} arising from non-local constitutive relations rather than from a quadratic tensor action with somewhat ad hoc non-local form factors.

Since the oscillations occur in both higher-derivative and infinite-derivative classes of theories, quite independently from one another, we hope that the above observations may prove helpful in extracting observational criteria on the gravitational potential at mesoscopic distances.

\section*{Acknowledgements}
The author benefited from discussions with Valeri P.\ Frolov, Hennadii Yerzhakov, and Andrei Zelnikov (all Edmonton), as well as Breno Loureiro Giacchini (Rio de Janeiro), and is moreover grateful for a Vanier Canada Graduate Scholarship administered by the Natural Sciences and Engineering Research Council of Canada as well as for the Golden Bell Jar Graduate Scholarship in Physics by the University of Alberta.

\begin{singlespace}

\end{singlespace}

\end{document}